# Binary nanocrystalline alloys with strong glass forming interfacial regions: Complexion stability, segregation competition, and diffusion pathways

Tianjiao Lei [1], Esther C. Hessong [1], Daniel S. Gianola [2], Timothy J. Rupert [1,3,*]

[1] Department of Materials Science and Engineering, University of California, Irvine, CA 92697, USA

[2] Materials Department, University of California, Santa Barbara, CA 93106, USA

[3] Department of Mechanical and Aerospace Engineering, University of California, Irvine, CA 92697, USA

* Email: trupert@uci.edu

## Abstract

Stabilization of grain structure is important for nanocrystalline alloys, and grain boundary segregation is a common approach to restrict coarsening. Doping can alter grain boundary structure, with high temperature states such as amorphous complexions being particularly promising for stabilization. Dopant enrichment at grain boundaries may also result in precipitate formation, giving rise to dopant partitioning between these two types of features. The present study elucidates the effect of dopant choice on the retention of amorphous complexions and the stabilization of grain size due to various forms of interfacial segregation in three binary nanocrystalline Al-rich systems, Al-Mg, Al-Ni, and Al-Y as investigated in detail using transmission electron microscopy. Amorphous complexions were retained in Al-Y even for very slow cooling conditions, suggesting that Y is the most efficient complexion stabilizer. Moreover, this system exhibited the highest number density of nanorod precipitates, reinforcing a recently observed correlation between amorphous complexions and grain boundary precipitation events. The dopant concentration at the grain boundaries in Al-Y is lower than in the other two systems, although enrichment compared to the matrix is similar, while secondary segregation to nanorod precipitate edges is much stronger in Al-Y than in Al-Mg and Al-Ni. Y is generally observed to be an efficient doping additive, as it stabilizes amorphous features and nanorod precipitates, and leaves very few atoms trapped in the matrix. As a result, all grains in Al-Y remained nanosized whereas abnormal grain growth occurred in the Al-Mg and Al-Ni alloys. The present study demonstrates nanocrystalline stability via simple alloy formulations and fewer dopant elements, which further encourage the usage of bulk nanostructured materials.

# 1. Introduction

A popular materials design strategy capable of achieving a synergy between high strength and good ductility is the development of heterogeneous microstructures [1,2,3]. For pure metals, common approaches to create non-uniform microstructures include mixing regions with dramatically different grain sizes, so that a gradient of plastic strain can build up during deformation to enhance strain hardening [4]. When dopant elements are added into the system, additional microstructural features that enhance mechanical properties may form, such as grain boundary segregation and precipitate formation. Importantly, these features can stabilize the microstructure up to high processing and operating temperatures. For example, Du et al. [5] examined the thermal stability of an Al-Mg-Sc-Zr-Mn system with Ca. After aging at 300 °C (a homologous temperature, $T_H \sim 0.61$) for 4 h and then 500 °C ($T_H \sim 0.83$) for 36 h, the grain size only increased a small amount, from $32.0 \pm 1.5$ μm in the as-cast condition to $33.9 \pm 2.3$ μm. The excellent thermal stability was attributed to both pinning effect of $Al_3(Sc,Zr)$ and good thermal resistance of $Al_4Ca$ particles at grain boundaries. Microstructure stabilization is especially important for nanostructured materials since their properties are closely associated with nanosized grains. In our previous study [6], nanocrystalline ternary Al alloys with a hierarchical microstructure containing grain boundary segregation, nanometer thick amorphous grain boundary complexions, nanoscale grain boundary precipitates, and large intermetallic particles with sizes ranging from sub-micrometer to a few micrometers were fabricated. The various types of reinforcements concurrently contributed to yield strengths higher than 900 MPa and stable plastic flow [7]. Moreover, the matrix grain sizes remained ~50 nm even after a hot-pressing temperature at 92% of the melting temperature of pure Al.

Amorphous complexions are characterized as grain boundary states with a disordered structure and uniform thickness [8,9,10]. These amorphous complexions can significantly stabilize the grain structure due to their lower free energy at high temperatures, and therefore the driving force for grain growth will be reduced [11,12,13]. For example, Schuler et al. [14] found that amorphous complexions enabled a new high-temperature grain size stability regime for a nanocrystalline Ni-W film fabricated using pulsed electrodeposition. The grain size increased from the as-deposited value of 35 nm to ~90 nm after being annealed at 900 °C ($T_H \sim 0.68$) for 1 h. However, grain growth to only ~55 nm was observed after annealing at a higher temperature of 1100 °C ($T_H \sim 0.79$) for the same time, which could be attributed to the onset of amorphous complexion formation at this higher temperature. In order to retain these features at room temperature, fast cooling is often used as a final processing step so that the complexions do not have enough time to transform back to an ordered state that is more stable at low temperatures. Despite this requirement, the ternary Al alloys (Al-Mg-Y, Al-Fe-Y, and Al-Ni-Y) studied in our previous work [6,7] all exhibited amorphous complexions even after a very slow cooling rate less than 1 °C/s, pointing to the excellent complexion stability within these systems. Moreover, the complexion thickness was similar across the three alloys even though they possessed different mixing enthalpies and atomic size mismatches, which have been predicted to mediate complexion formation [15]. Therefore, a dominant alloy selection consideration should exist for the formation of thick and stable amorphous complexions, yet the chemical complexity of the ternary systems makes it challenging to isolate the effect of each dopant element.

In addition to complexion transitions, heavily doped grain boundaries may be pushed into a thermodynamic regime where precipitation of bulk secondary phases occurs. Since both precipitates and grain boundaries are typically enriched with dopant atoms, segregation competition may then occur. For compositionally complex alloys, the segregation sequence of various types of elements to grain boundaries can result in complicated interfacial precipitation behavior. For example, Mantha et al. [16] performed in situ transmission electron microscopy heating experiments to investigate the phase evolution within a nanocrystalline CoCuFeMnNi alloy processed by high pressure torsion. These authors observed competition between grain boundary segregation and depletion, which led to precipitate formation that was dependent on the heating temperature. For instance, from 250-300 $^\circ$C, Co, Ni and Cu segregated to grain boundaries while Fe and Mn depleted at grain boundaries, and no precipitates formed. As temperature raised to 340 $^\circ$C, the segregation of Fe and Co resulted in formation of B2 precipitates at triple junctions. In our previous nanocrystalline Al-Mg/Ni/Fe-Y systems, the formation of nanorod carbide precipitates with a core-shell structure were observed to nucleate from amorphous complexion sites [6,7]. Both the interfaces and precipitate shells were enriched with the same dopant elements, and the precipitates shared similar growth kinetics with the matrix grains during isothermal annealing treatments [17]. Thus, a detailed picture of the interplay between segregation competition at interfaces, precipitate chemistry, and grain size stabilization remains unclear.

In the present study, in order to investigate the effect of dopant element on the formation and retention of amorphous complexions and study the effect of various interfacial features on stabilizing the microstructure, three binary alloy systems (Al-Mg, Al-Ni, and Al-Y) that cover a wide range of mixing enthalpies, segregation enthalpies, and atomic size mismatch, were carefully examined. For each alloy, samples were either slow cooled down to room temperature or annealed followed by rapid water quenching, in order to probe complexion stability. Amorphous complexions were retained in Al-Y for both cooling conditions, suggesting that Y provides the highest stabilization of the amorphous complexions as rapid cooling is not needed. In addition, this system exhibits a larger number density of precipitates than the other two alloys, pointing to the strong correlation between complexion stability and precipitate formation. Elemental quantification revealed that most of the Y atoms segregated to either grain boundaries or precipitate edges, with very few left in the matrix in solution. The Al-Y alloy also had the highest thermal stability of the three materials studied here, as the grain sizes were below 100 nm for all conditions and no signs of abnormal grain growth were observed. Al-Ni and Al-Mg demonstrated decreasing levels of complexion and grain size stabilization. The present study provides important information on amorphous complexion stability, segregation competition between interfacial features, and diffusion pathways of dopant atoms, all of which shed light on efficient strategies for designing stable heterogeneous nanostructured materials.

## 2. Materials and methods

To fabricate Al-Mg, Al-Ni, and Al-Y alloys, powders of elemental Al (Alfa Aesar, 99.97%, -100+325 mesh), Mg (Alfa Aesar, 99.8%, -325 mesh) or Ni (Alfa Aesar, 99.9%, APS 2.2-3.0 micron) or Y (Alfa Aesar, 99.6%, -40 mesh) were first ball milled for 10 h in a SPEX SamplePrep 8000M high-energy ball mill. For all three alloys, the nominal dopant concentration was 2 at.%. Our prior study of a similar Al-Ni-Y ternary alloy [17] showed that overall alloy composition after ball milling matched the nominal concentration of powders added. As such, only the nominal concentration will be quoted in this work. A hardened steel vial and milling media were used with a ball-to-powder weight ratio of 10:1, while 3 wt.% stearic acid ($C_{18}H_{36}O_2$) was used as a process control agent to prevent excessive cold welding. We note that the stearic acid provides a potential source of C, H, and O impurities, which may result in formation of secondary phases in the final materials. The milling process was conducted in a glovebox filled with Ar gas at an $O_2$ level <0.05 ppm to avoid oxidation.  After milling, the alloyed powders were transferred into a ~14 mm inner diameter graphite die set, and then consolidated into cylindrical bulk pellets using an MTI Corporation OTF-1200X-VHP3 hot press consisting of a vertical tube furnace with a vacuum-sealed quartz tube and a hydraulic press.  The powders were first cold pressed for 10 min under 100 MPa at room temperature to form a green body, and then hot pressed for 1 h under 100 MPa at 585 °C.  The hot-pressing temperature was approximately 92% of the melting temperature of pure aluminum.  The heating rate used to reach the target temperature was 10 °C/min.  Followed hot pressing, the pellets were slowly cooled down to room temperature with a cooling rate less than 1 °C/s by turning off the furnace.  Subsequently, one specimen of each alloy was annealed at 585 °C for 10 min and then quenched into a water bath to achieve a much higher cooling rate. Therefore, for each alloy system, two types of samples were fabricated: (1) a specimen that was naturally cooled to room temperature with a slow cooling rate after hot pressing, referred to as *naturally cooled*, and (2) a specimen that was annealed for a short time (10 min) and then rapidly water quenched following the consolidation process, referred to as *fired + quenched*.

The consolidated cylindrical pellets were polished with SiC grinding paper down to 1200 grit and then with monocrystalline diamond pastes down to 0.25 μm prior to microstructural characterization.  X-ray diffraction (XRD) measurements were conducted using a Rigaku Ultima III X-ray diffractometer with a Cu Kα radiation source operated at 40 kV and 30 mA and a one-dimensional D/teX Ultra detector.  Phase identification and fraction were obtained using an integrated powder X-ray analysis software package (Rigaku PDXL).  Scanning/transmission electron microscopy (S/TEM) was used to examine grain sizes and precipitate morphology in a JEOL JEM-2800 S/TEM operated at 200 kV.  High-resolution TEM was employed to characterize amorphous grain boundary complexions using the same microscope.  The elemental distribution in the vicinity of grain boundaries and precipitates was examined using high-angle annular dark field (HAADF)-STEM combined with energy-dispersive spectroscopy (EDS) in a JEOL JEM-ARM300F Grand ARM TEM operated at 300 kV with double Cs correctors and dual 100 mm$^2$ silicon drift detectors. For EDS mapping, a probe current of 204 pA was used, and each precipitate edge/grain boundary was carefully tilted to an edge-on condition to ensure accurate quantification of local composition. All TEM samples were fabricated using the FIB lift-out method [18] with a Ga$^+$ ion beam in the FEI Quanta 3D FEG dual-beam SEM/FIB microscope equipped with an OmniProbe. A final polish at 5 kV and 48 pA was used to minimize the ion beam damage to the TEM sample.

## 3. Results and Discussion

### 3.1. Microstructural Features and Grain Size Stabilization

XRD was first employed to examine the overall phase content and average microstructures across all alloys and processing conditions, and scans corresponding to Al-Mg, Al-Ni, and Al-Y are presented in Figures 1(a1)-(c1). All systems exhibited peaks corresponding to both a face-centered cubic Al and a trigonal Al$_4$C$_3$ phase. The formation of the carbide phase was due to C atoms introduced during the ball milling process, specifically from the stearic acid (C$_{18}$H$_{36}$O$_2$) used to prevent cold welding and powder agglomeration. Besides aluminum carbides, intermetallic phases formed in the Al-Ni and Al-Y alloys, which were identified as orthorhombic Al$_3$Ni and trigonal Al$_3$Y, respectively. For Al-Mg, there was no intermetallic formation, most likely due to the much higher equilibrium solid solubility of Mg in Al [19]. Comparing the two cooling conditions, no obvious difference in the intensity, position, and width of the peaks was observed, pointing to similar phase fractions and the retention of nanocrystalline grain structures. It is worth noting that although ball milling may result in amorphous phase formation in some alloy systems [20,21,22], no diffuse peak corresponding to amorphous phases was observed in XRD scans of the as-milled powders in the present study. All samples were nanocrystalline after the ball milling process. Table I lists the weight percent of secondary phases obtained using a whole powder pattern fitting on the XRD data, where the *naturally cooled* and *fired + quenched* specimens show similar values for $Al_4C_3$ (~8 wt.%) in Al-Mg, and for $Al_3Y$ (~8 wt.%) and $Al_4C_3$ (~10 wt.%) in Al-Y. For Al-Ni, the intermetallic phase fraction did not change much (~10 wt.%), while the carbide weight percent increased from 2.5% to 9.2% after the annealing and quenching treatment.

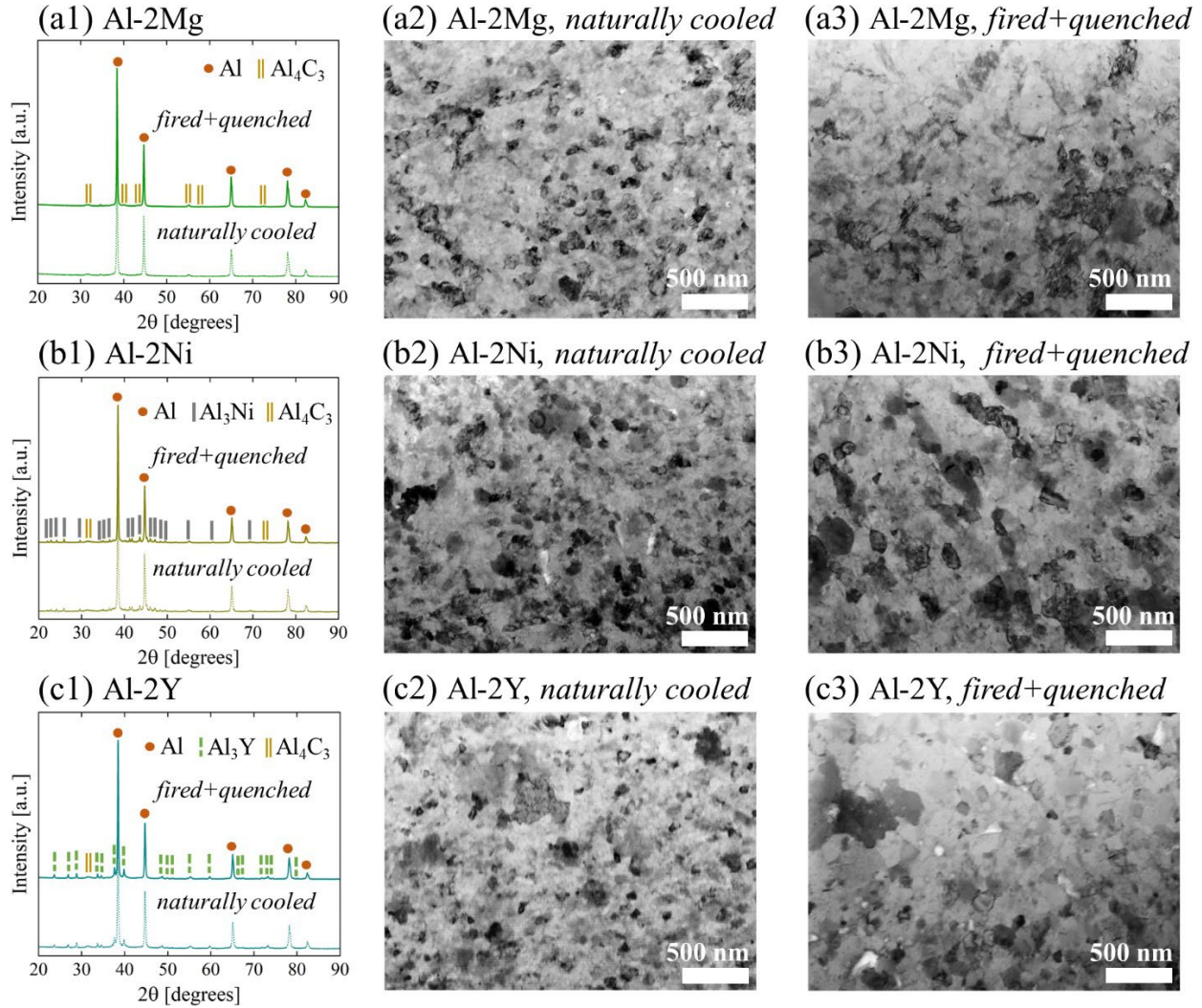

**Figure 1.** (a1-c1) XRD scans of Al-Mg, Al-Ni, and Al-Y, respectively, with each alloy being either *naturally cooled* after consolidation or fired followed by water quenching (*labelled as fired + quenched*). (a2-c2) BF-STEM images showing grain sizes and morphology in the *naturally cooled samples*. (a3-c3) BF-STEM images correspond to the *fired + quenched* condition, where grain sizes slightly increased while the grain morphology remained the same.

**Table I.** Weight percent of secondary phases in all sample conditions.

| Alloy | Secondary phase | Cooling condition | Weight percent (%) |
|---|---|---|---|
| Al-2Mg | $Al_4C_3$ | *naturally cooled* | 8.3 |
| | | *fired + quenched* | 8.6 |
| Al-2Ni | $Al_4C_3$ | *naturally cooled* | 2.5 |
| | | *fired + quenched* | 9.2 |
| | $Al_3Ni$ | *naturally cooled* | 9.8 |
| | | *fired + quenched* | 10.3 |
| Al-2Y | $Al_4C_3$ | *naturally cooled* | 9.5 |
| | | *fired + quenched* | 10.2 |
| | $Al_3Y$ | *naturally cooled* | 8.0 |
| | | *fired + quenched* | 8.1 |

Grain sizes were next examined, and the Al matrix sizes obtained from the XRD data point to slight coarsening during the *fired + quenched* treatment. Grain sizes evolved from an average size of 64 nm to 89 nm for Al-Mg, 55 nm to 98 nm for Al-Ni, and 50 nm to 88 nm for Al-Y. In order to verify the grain growth, TEM investigations were employed and confirmed that the Al matrix grain sizes increased by a small extent from *naturally cooled* (Figures 1(a2)-(c2)) to *fired + quenched* (Figures 1(a3)-(c3)) conditions. The grain morphology remained equiaxed in all samples originated from the mechanical milling and consolidation process. However, further examination at lower magnifications revealed that several matrix grains coarsened to a few micrometers in Al-Mg and Al-Ni, with representative images corresponding to the *fired + quenched* condition for all three alloys presented in Figure 2 and abnormal grains marked with yellow arrows. A larger region of the material was composed of abnormally grown grains in the Al-Mg alloy as compared to the Al-Ni alloy. These large grains do not substantially contribute to the XRD peak broadening because this technique only yields average information [23], and the micrometer grain size is well beyond the detection limit [24]. For Al-Y, all matrix grains were below 100 nm (Figure 2(c)) and no abnormal grain growth was observed,

(a) Al-2Mg, *fired+quenched*   (b) Al-2Ni, *fired+quenched*   (c) Al-2Y, *fired+quenched*

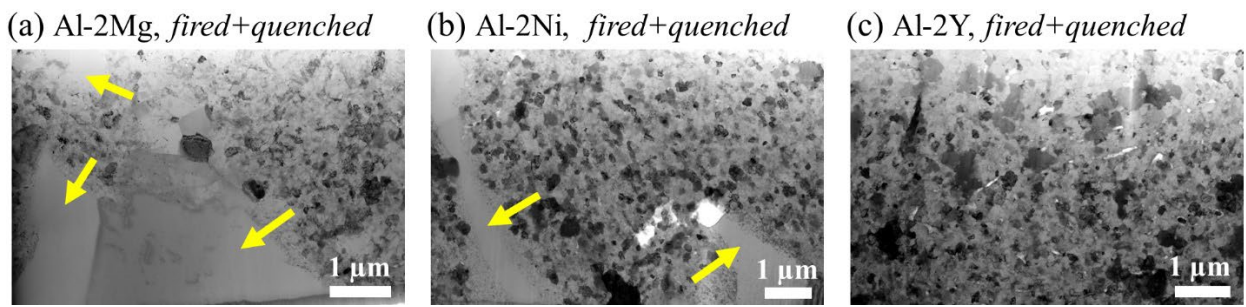

**Figure 2**. (a)-(c) Low-mag BF-STEM micrographs corresponding the *fired + quenched* Al-Mg, Al-Ni, and Al-Y, respectively. The grains remained nanosized only in the Al-Y system, while abnormal grain growth existed in both Al-Mg and Al-Ni.

demonstrating that Y can stabilize the microstructure more effectively than Mg and Ni. In order to understand the various segregation tendencies and stabilizing abilities of Mg, Ni, and Y in these nanostructured alloys, local microstructural features are examined in detail below.

Since nanostructured alloys contain a large volume fraction of interfaces, interfacial structure and chemistry significantly affect the stability of the grain structure. Among the various types of interfaces, amorphous grain boundary complexions, which usually begin to form at 0.6-0.85$T_H$ [25] and are stable at high temperatures, have been shown to dramatically improve the thermal stability of nanocrystalline alloys. Because the hot pressing in the present study was performed at $T_H$ = 0.92, higher than the formation temperature for amorphous complexions, high-resolution TEM was used to verify the existence of these features. For each sample, at least ten grain boundaries were examined. Figures 3(a)-(c) show representative grain boundary structures complexions from the Al-Mg, Al-Ni, and Al-Y alloys, respectively. For the *naturally cooled* condition (top panels), only ordered grain boundaries were observed in all examined area for Al-Mg and Al-Ni. In contrast, amorphous complexions were retained in this condition in Al-Y, with an example outlined by dashed lines in the top panel of Figure 3(c) having a thickness of 2.8 nm. Prior to the measurement of complexion thickness, the defocus value during high resolution TEM imaging was cycled between negative and positive values to identify interfaces that were amorphous, as detailed in Ref. [26]. After the annealing and quenching process, which is meant to attempt to freeze in high temperature structures, Al-Mg still only exhibited ordered boundaries (lower panel of Figure 3(a)). However, amorphous complexions emerged in Al-Ni (lower panel of Figure 3(b)) due to the much higher cooling rate used in this process. In addition, much thicker amorphous complexions were observed in Al-Y after firing and quenching, with a representative thickness of 7.2 nm shown in the lower panel of Figure 3(c), demonstrating that retained complexion thickness is a function of cooling rate. It is worth noting that the complexion thicknesses observed here for Al-Y (>7 nm) were similar to those found in a ternary Al-Ni-Y alloy (7-9 nm) with similar heat treatment in our previous study [6]. Comparing binary and ternary systems, the introduction of a second dopant species can reduce the formation energy for amorphous complexions and therefore give rise to larger complexion thicknesses in some cases. For example, Grigorian and Rupert [26] quantitively investigated the distribution of complexion thickness in Cu-Hf, Cu-Zr, and Cu-Zr-Hf alloys, and observed that the ternary system exhibited much thicker complexion (with the highest value being ~9 nm) than the two binary samples (the largest thicknesses of which were less than 6 nm). In contrast, the similar complexion thickness of Al-Y in the present study with the observations for Al-Ni-Y in Ref. [6] points to the dominant effect of Y on the nucleation and stabilization of amorphous complexions in Al-based alloys.

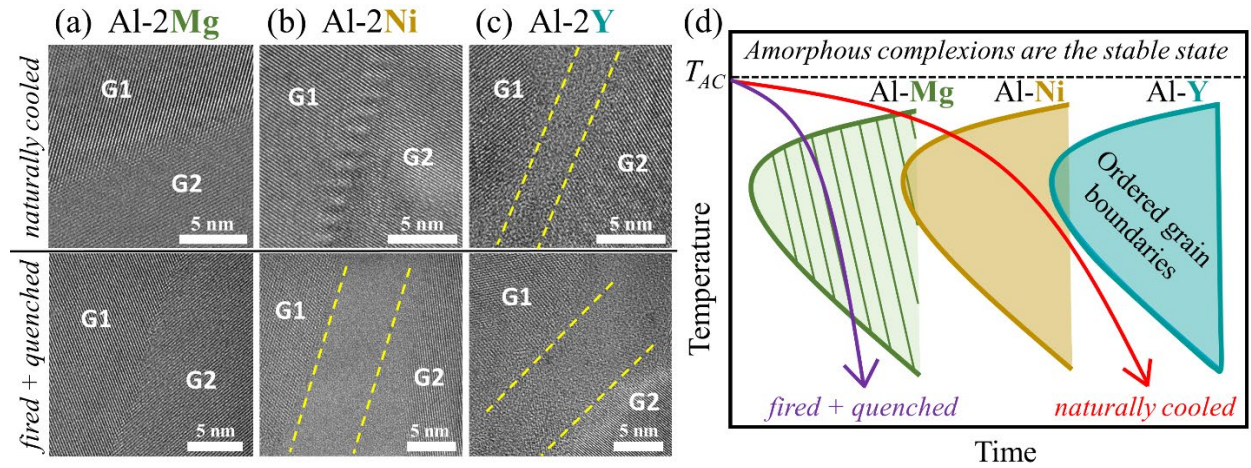

**Figure 3.** (a-c) High-resolution TEM micrographs of grain boundary structure for *naturally cooled* (top row) and *fired + quenched* (bottom row) Al-Mg, Al-Ni, and Al-Y, respectively, where amorphous complexions are enclosed between dashed lines. (d) Schematic illustration of time-temperature-transformation (TTT) curves for the three alloy systems where the time for order-to-disordered complexion transition is fastest for Al-Mg, which is why no amorphous complexions were observed, and slowest for Al-Y, which is why even very slow cooling results in amorphous complexions.

Prior work [15] has suggested that amorphous complexions can form if the grain boundary reaches a local composition that promotes high glass forming ability, with comparisons to bulk metallic glasses being useful. For example, Al-Ni-Y is a well-known glass forming system, and fully amorphous $Al_{86}Ni_8Y_6$ (at.%) rod samples with a diameter of 1 mm have been fabricated by casting molten master alloys into a copper mold [27]. In this example from the literature, the composition was selected based on simulation results [28] that showed atomic ratios of 16.9 for Al:Y and 9.4 for Al:Ni can introduce energetically favored solute-centered clusters into the system so that the glass forming ability will be significantly improved. Therefore, the ternary Al-Ni-Y system would be hypothesized to be a promising choice for amorphous complexions as well, which was confirmed by our previous studies [6,17]. Literature data does not exist for all of the binary systems studied in the present study, but a few examples allow for unique aspects of grain boundary amorphization to be discussed. Louzguine-Luzgin et al. [29] obtained amorphous Al-Y thin ribbons prepared by melt spinning despite the existence of multiple crystalline intermetallic phases predicted from the equilibrium phase diagram for this alloy. Therefore, this binary system should have reasonable glass forming ability and this explains its ability to form amorphous complexions. For Al-Ni, Sheng et al. [28] also employed melt spinning and hardly observed any glassy phase from X-ray diffraction measurements. The reduced glass forming ability of Al-Ni is therefore consistent with the fact that rapid cooling is required to capture amorphous complexions in the nanocrystalline Al-Ni alloys studied in this work. The reason that glass forming ability can be used to describe both metallic glasses and amorphous complexions is that these features form by either bulk melting or premelting transitions. Due to preexisting structural disorder at the grain boundaries, the transition from ordered to amorphous complexions occurs at a temperature that is below the bulk melting temperature. Although amorphous complexions are in equilibrium at high temperatures, they are not stable at room temperature and still need to be quenched into the microstructure, just like bulk metallic glasses must be quenched. However, the fact that the quenching rate can be very slow for some binary alloy systems and amorphous complexions are still retained (e.g., Al-Y) demonstrates the suppression of a transition back to the ordered state.

It should be noted that although grain boundary complexions can undergo transitions similar to phase transformations of bulk materials, these complexions are interfacial states that cannot exist independently and require compatibility with the grains on either side to find a local equilibrium configuration [9,10]. The confinement provided by the adjacent grains actually facilitates the formation of amorphous complexions, as the limited thickness of a grain boundary region (typically on the order of a nanometer thick) compared to the dimensions of a bulk material means that disordering only must occur locally, making the transition to a disordered state easier. Other factors, e.g., the fact that a random high angle grain boundary is already more disordered than a highly symmetric boundary, may also play a role.

Comparing the three dopant element species (Mg, Ni, and Y), amorphous complexions were observed only in Al-Y at low cooling rates (less than 1 ºC/s), suggesting that much higher cooling rates are needed to kinetically freeze in these features in Al-Mg and Al-Ni. Secondary evidence of this point is provided by the larger complexion thickness in Al-Y than in Al-Ni, with both alloys more effective complexion formers than Al-Mg (with a complete lack of amorphous complexions) for the *fired + quenched* condition (lower panel of Figures 3(a), (b), and (c)). Based on these observations, a schematic time-temperature-transformation (TTT) of the grain boundary structural transition for the three alloys can be constructed (Figure 3(d)). The transition from disordered complexion to ordered grain boundary state occurs very quickly in Al-Mg and shifts to longer time as the dopant changes from Mg→Ni→Y, with the highest complexion stability in Al-Y.  By examining grain boundary structures in Cu-rich alloys, Schuler and Rupert [15] proposed a set of selection rules for amorphous complexion formation, including negative pair-wise mixing enthalpy and large atomic size mismatch.  In the present study, the mixing enthalpies are -38 kJ/mol for Al-Y, -22 kJ/mol for Al-Ni, and -2 kJ/mol for Al-Mg [30], all of which are negative and therefore satisfy one of the criteria in Ref. [15] to different degrees.  For the difference in atomic sizes, Al-Y exhibits the highest value (29%) compared to Al-Mg (14%) and Al-Ni (10%) [31].  While both metrics would push all three alloys in the right direction to host amorphous complexions, the relative importance of each can be isolated by comparing against our experimental observations.  If solely based on the atomic size mismatch criterion, Al-Mg would be better than Al-Ni at forming amorphous complexions.  This is not the case and instead the TEM data aligns with a hypothesis that mixing enthalpy is the most important parameter for determining amorphous complexion stability.  The fact that no complexions were observed in Al-Mg suggests that a negative mixing enthalpy of only 2 kJ/mol, while technically being negative and preferring unlike bonds, is not enough to allow amorphous interfaces to be retained.  For bulk metallic glasses, Takeuchi and Inoue [32] proposed that alloy systems require at least a mixing enthalpy of -15 kJ/mol to achieve high glass forming ability by quantitatively investigating the effect of both atomic size mismatch and mixing enthalpies on critical cooling rates of a variety of systems.  This similarity in the existence of critical cutoff values of mixing enthalpies provides another connection between amorphous complexions and bulk metallic glasses.

In addition to their own stabilization effect on the grain structure, amorphous complexions can give rise to other interfacial features which further enhance microstructural stability [6,7,17].

Therefore, HAADF-STEM was employed to examine precipitates within each alloy, with Figures 4(a)-(c) showing representative micrographs for *fired + quenched* Al-Mg, Al-Ni, and Al-Y, respectively. In all systems, rod-shape precipitates are formed at grain boundary sites, as indicated by red arrows. The precipitate interior shows similar contrast as the matrix, suggesting that the interior is mainly composed of Al and is identified as the $Al_4C_3$ phase from the XRD scans (confirmed with EDS below). The edges of these precipitates exhibit a different contrast from the precipitate interior and the matrix, pointing to a change in local concentration. In addition, grain boundaries possess similar contrast to the precipitate edges, suggesting that both features experience dopant segregation behavior, which is investigated in more detail in the following sections. Besides the nanoscale interfacial features, larger regions that are much brighter than the matrix were observed in the Al-Ni and Al-Y alloys (as indicated by yellow arrows), which correspond to $Al_3Ni$ and $Al_3Y$, respectively. For the remainder of this paper, the $Al_4C_3$ phase will be referred to as nanorod precipitates and the $Al_3Ni/Al_3Y$ phases will be referred to as intermetallic particles.

Although the carbide precipitates existed in all three alloy systems, their amount and morphology varied with the chosen dopant element. In order to obtain the number density, at least six micrographs were evaluated for each alloy, with the average values included in Figures 4(a)-(c). The number densities of the nanorods in Al-Mg ($3185/\mu m^3$) and in Al-Ni ($5166/\mu m^3$) are much lower than that in Al-Y ($13760/\mu m^3$). In addition, the precipitates in the Al-Ni alloy were noticeably wider than those in either Al-Mg or Al-Y, appearing less elongated in Figure 4(b). To quantitively compare the precipitate size, more than 100 precipitates were measured for each system, and the corresponding cumulative distribution functions are shown in Figures 4(d) and (e) for nanorod precipitate length and width, respectively. For the length, the curve of Al-Mg appears at the highest values, meaning that while the nanorods are fewer in number in this alloy, they have the longest dimensions. Consequently, the average precipitate length in Al-Mg (37.7 ± 20.0 nm) is larger than in the other two systems, which have similar values (29.0 ± 10.5 nm for Al-Ni and 29.7 ± 16.4 nm for Al-Y). The length distribution for fractions < 0.7 in Al-Ni and Al-Y are similar, suggesting that only the longest nanorods in the alloys demonstrate a systematic difference, with the Al-Y nanorods being up to 100 nm long while those in Al-Ni stay below ~70 nm. For the precipitate width, the distribution shifts to larger value as the dopant element changes from Y → Mg → Ni, and the average widths in Al-Y, Al-Mg, Al-Ni are 7.4 ± 2.4 nm, 10.2 ± 3.5 nm, and 15.1 ± 3.9 nm, respectively. The combinations of widths and lengths for the nanorods result in dramatically different overall shapes in the three systems. The length-to-width aspect ratio of the nanorods is 4.0 for Al-Y and 3.7 for Al-Mg, roughly twice the value of 1.9 measured for Al-Ni.

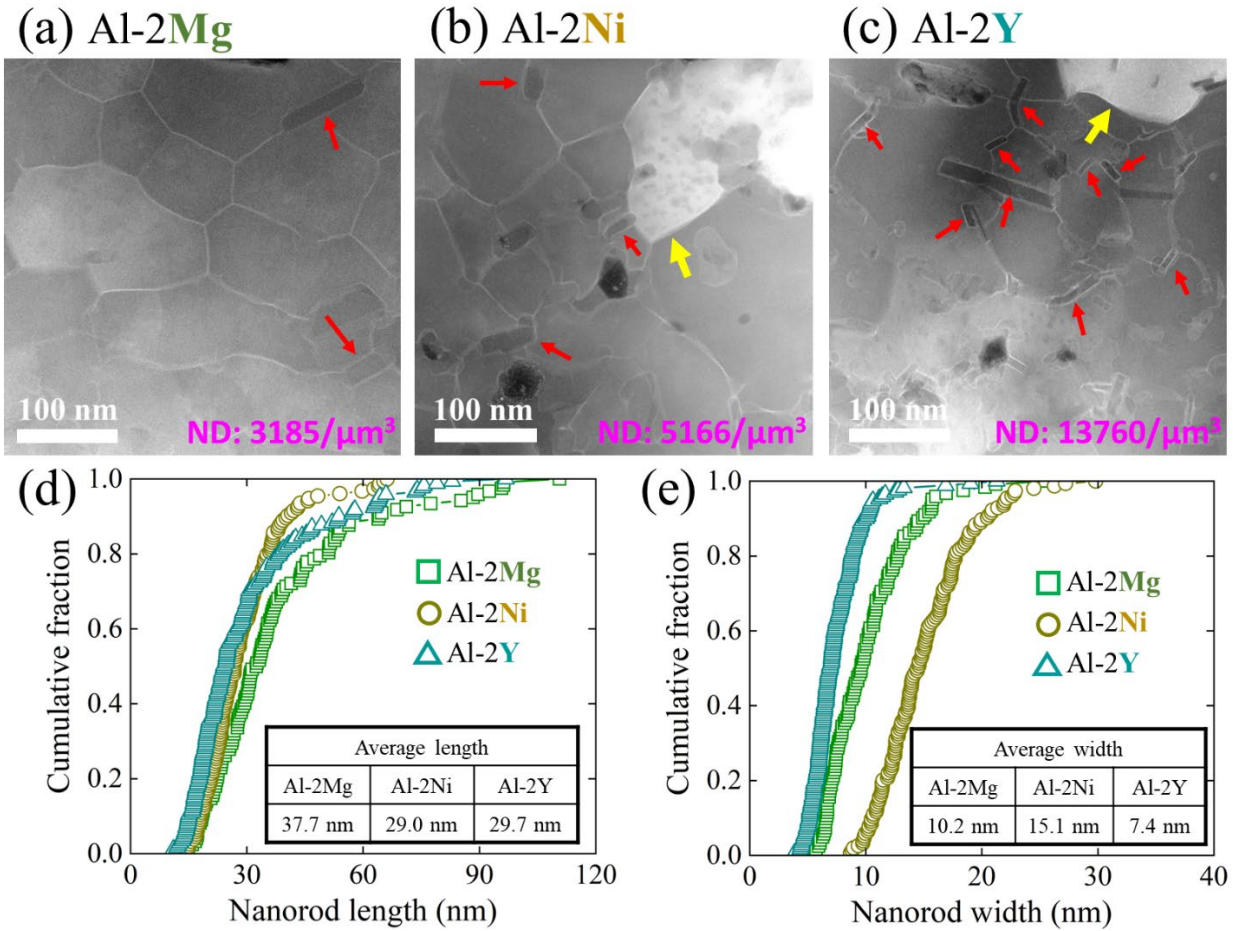

**Figure 4.** Representative HAADF-STEM micrographs showing morphology, location, and number density (labelled as "ND" and the values are shown) of carbide nanorod precipitates (indicated by red arrows) for *fired + quenched* (a) Al-Mg, (b) Al-Ni, and (c) Al-Y. Intermetallic phases in Al-Ni and Al-Y are denoted by large yellow arrows. (d) and (e) correspond to cumulative distribution functions of the precipitate length and width, respectively, and the average values for all system are also presented in the plot.

Considering the data in Figures 3 and 4 together shows that the Al-Y alloy, which is more effective at sustaining amorphous complexions, also exhibits a higher number density of nanorod precipitates, while the Al-Mg system with no retained amorphous complexions shows by far the fewest precipitates. The complexion thickness as well as the precipitate number density in the Al-Ni alloy fall between those of the other two alloys. This observation across various alloy systems provides clear evidence that a correlation exists between the presence of amorphous complexions and nanorod precipitate density. Garg et al. [33] observed that thicker complexion distributions correlate with higher complexion volume fractions when modeling a polycrystalline grain boundary network with hybrid atomistic Monte Carlo/molecular dynamics simulations. Consequently, the thicker amorphous complexions in Al-Y (Figure 3(c)) are an indication that there should be a larger number fraction of grain boundaries that have transformed into complexions in this alloy and, therefore, more nucleation sites for precipitates. It is worth noting that the presence of carbide nanorods in Al-Mg suggests that amorphous complexions were also present in this system at the high temperatures used for hot pressing and annealing, since amorphous complexions were previously found to be the only nucleation site for precipitates. However, the amorphous complexions in Al-Mg were never observed for either cooling conditions, meaning that even a water quenching treatment was not fast enough to freeze them in and these features must have quickly transformed back to an ordered state. This behavior is shown by the green curve in Figure 3(d). Therefore, although nanorod precipitates and amorphous complexions form within similar temperature ranges [6], the precipitates are much more stable against dissolution upon cooling than the complexions. As a whole, the presence of stable amorphous complexions and a high density of nanorod carbide precipitates in the Al-Y alloy are responsible for the improved resistance to abnormal grain growth. It is worth noting that if other processing techniques are used to fabricate nanocrystalline Al-rich alloys where no C is introduced, the stabilizing effect of the carbides should be lost. However, stabilization from amorphous complexions would be active. Since Y is the most efficient complexion stabilizer followed by Ni and then Mg, the stability of the three systems would likely still be ordered as Al-Y > Al-Ni > Al-Mg.

### 3.2. Segregation Competition and Dopant Diffusion Pathways

Since both nanorod precipitates and grain boundaries are enriched with dopants, the next step is to quantify the extent of the segregation behavior and identify competition for dopants. For grain boundary examinations, the boundaries were classified based on whether they were in the vicinity of a nanorod or not, to understand if nanorod growth depletes boundaries of dopants. Figures 5(a) and (b) show the local composition of two boundaries without nanorods in the *fired*

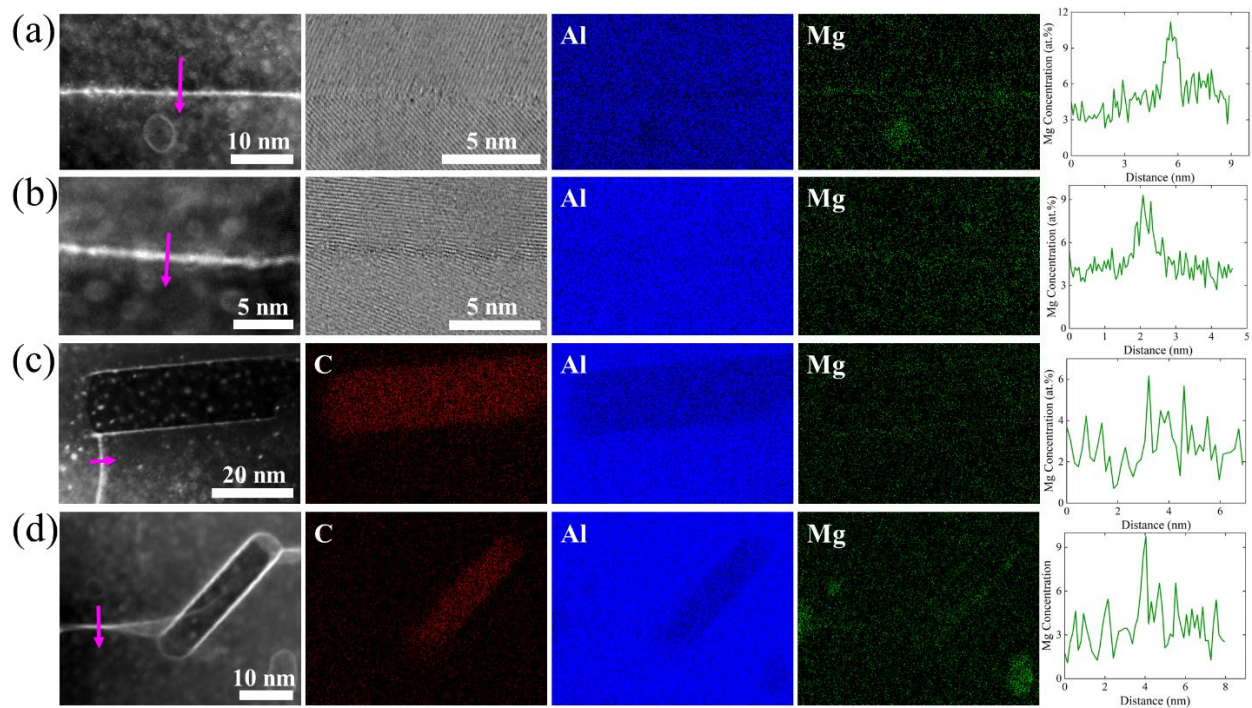

**Figure 5.** Grain boundary segregation in *fired + quenched* Al-Mg system. (a), (b) Two representative grain boundaries without nanorod precipitates in the vicinity, and (c), (d) two boundaries connected to nanorod precipitates.

*+ quenched* Al-Mg system. Care was taken to ensure an edge-on condition for all examined boundaries before EDS mapping, as verified by the high-resolution BF-STEM images in the second panel. The mapping of Mg demonstrates an enrichment of dopant atoms at the grain boundaries as the intensity is higher in the same region. The grain boundary segregation is further verified by the obvious peak in the Mg concentration from the line scans, from which the ratio of the peak concentration to the matrix concentration, termed the *enrichment factor*, is obtained. The enrichment factors are 2.5 and 2.2 for Figures 5(a) and (b), respectively. These concentration ratios are close to values reported in previous work on Al-Mg alloys. For instance, Malis and Chaturvedi [34] employed energy dispersive X-ray microanalysis to study grain boundary segregation in an Al-8 wt.% Mg alloy prepared by casting, cold rolling, and then solution heat treating followed by water quenching. The final microstructure consisted of equiaxed grains with an average diameter of 20 μm, and the Mg amount at grain boundaries was two to three times higher than the overall alloy composition. Liu and Adams [35] reported similar enrichment factors from Monte Carlo simulations as grain boundary concentrations ranged from 20-40 at.% Mg in an Al-10 at.% Mg alloy. When nanorods are nearby, as shown in Figures 5(c) and (d), Mg segregation to grain boundaries is still evident and enrichment factors of 1.9 and 3.1 were measured, respectively. Therefore, the formation of the nanorod precipitates does not seem to affect the grain boundary segregation. This is in contrast to a previous study on a nanocrystalline Al-5 at.% Mg alloy [36], where the dopant concentration dropped from 12 at.% to 2.5 at.% at grain boundaries after precipitate formation. However, the precipitates in Ref. [36] were an $Al_3Mg_2$ intermetallic phase rather than a carbide phase. Consequently, no Mg atoms are required for the core of the precipitates in the present study.

Figure 6 shows similar grain boundary concentration data, both away from and near nanorods, for the *fired + quenched* Al-Ni system. The elemental maps show clear Ni enrichment, which is confirmed by line scans across these boundaries as presented in the far-right column. The enrichment factors obtained from these line scans are 3.8 and 2.7 for the boundaries away from the nanorods and 4.0 and 2.8 for the boundaries nearby nanorods, again showing that nanorod formation and growth does not noticeably deplete dopant concentration at the grain boundaries. For the Al-Y system, the dopant concentrations at all grain boundaries (obtained from line scans in Figure 7) are much lower than those observed for the Al-Mg and Al-Ni alloys. However, there is also less Y in the grain interiors, due to intermetallic formation and a much higher number density of nanorod carbides with strongly doped precipitate edges. A back-of-the-envelope calculation suggests that the depletion of Y from both grain boundaries and grain interior can be explained by the level of segregation observed at the carbide nanorod edges and intermetallic phases (see Supplemental Materials Section 1 and Figure S1). Therefore, normalizing grain boundary concentration by the corresponding matrix dopant content gives enrichment factors that are similar to the other two systems. It is worth noting that most Y atoms either form $Al_3Y$ intermetallic particles or segregate to edges of nanorod precipitates, which will be shown in detail later. For the boundaries without nanorod precipitates, the enrichment factors are 2 and 2.9, while those for boundaries with nanorods are 2.3 and 2.4. Therefore, nanorod formation and growth does not noticeably deplete the grain boundary concentration for any of the alloys studied here. Comparing the three binary systems, although the absolute values of dopant concentrations at grain boundaries are much higher in Al-Mg and Al-Ni than in Al-Y, the dopant enrichment factors at grain boundaries are similar across all systems.

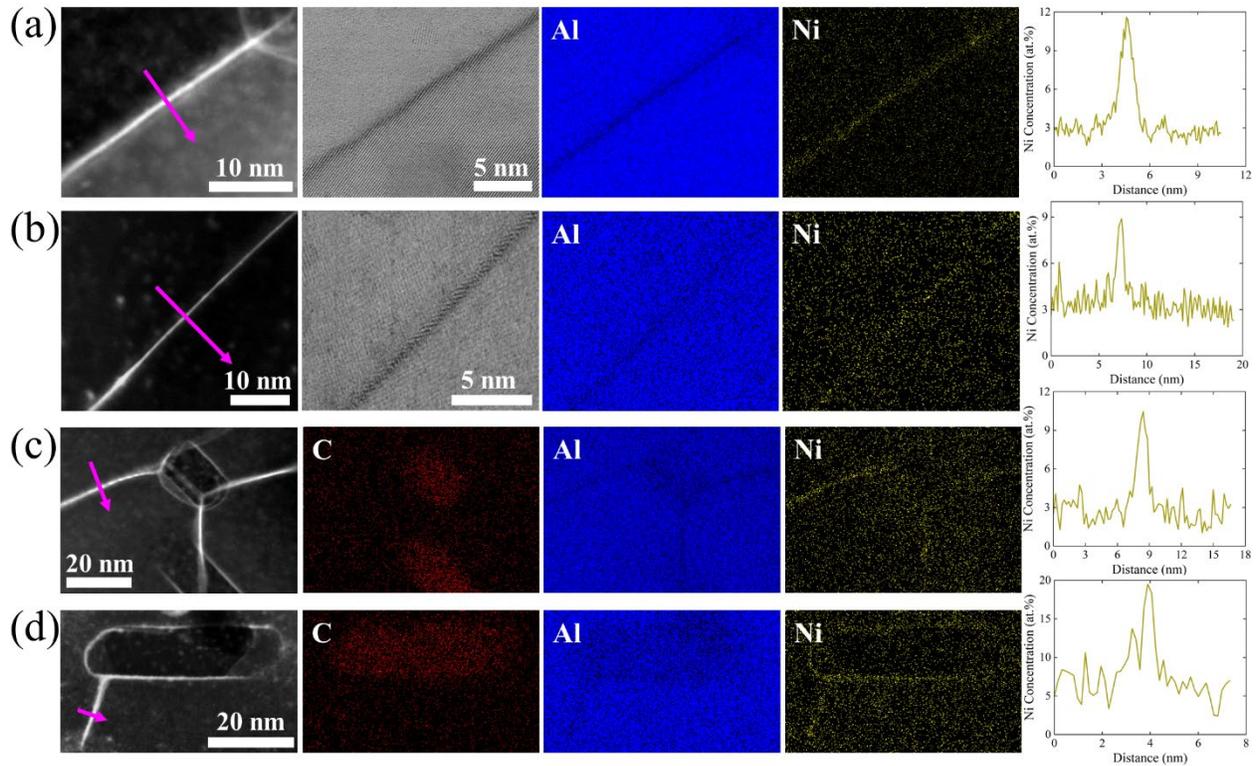

**Figure 6.** Grain boundary segregation in *fired + quenched* Al-Ni system. (a), (b) Two representative grain boundaries without nanorod precipitates in the vicinity, and (c), (d) boundaries connected to nanorod precipitates, where Ni atoms strongly segregated to all boundaries.

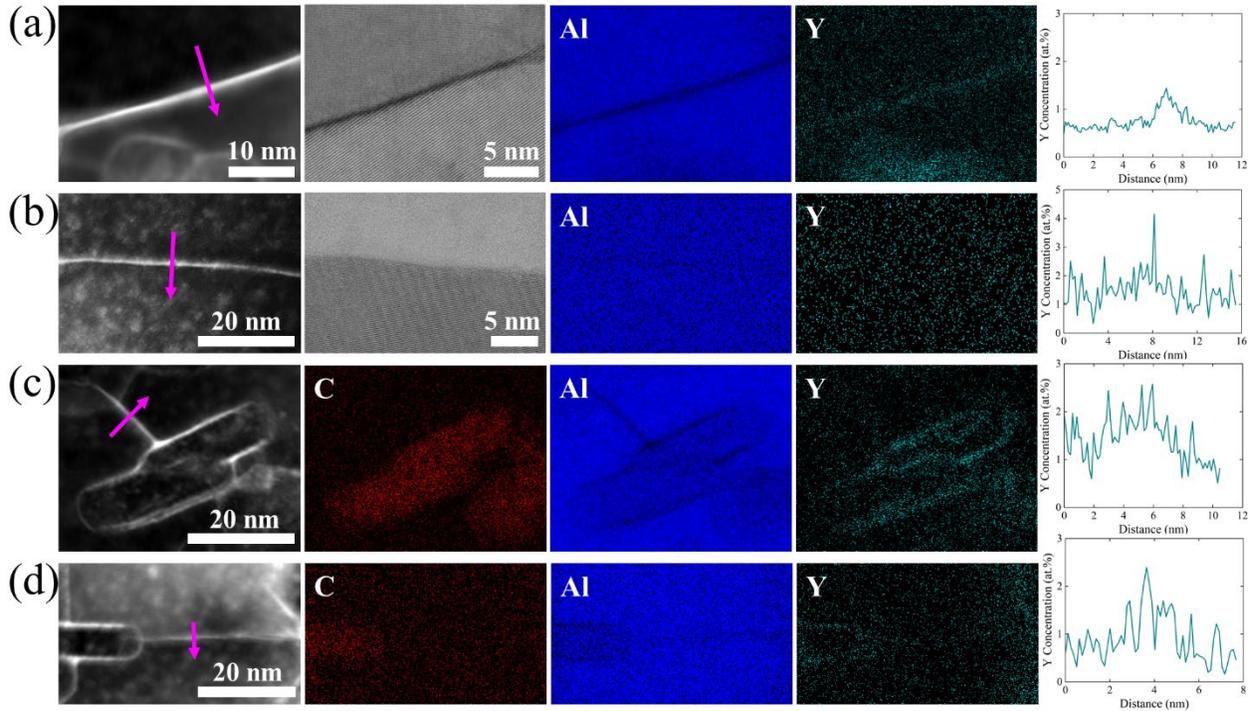

**Figure 7.** Grain boundary segregation in *fired + quenched* Al-Y system. (a), (b) Representative grain boundaries without nanorod precipitates in the vicinity, and (c), (d) boundaries connected to nanorod precipitates. Segregation of Y atoms was weak in all of the cases, especially for the boundaries connected to nanorod precipitates, where no obvious segregation was observed.

The segregation to nanorod edges is quantified next. Figure 8 shows the chemistry of representative nanorod precipitates in the *fired + quenched* condition for the three alloys, where the interior is composed of Al and C, confirming that these features are the Al₄C₃ phase. At the precipitate edge, dopant atoms segregated to the longer side much more strongly than to the shorter side, and Figure 9 presents a low-magnification view of these precipitates, which clearly shows the segregation difference. The anisotropic segregation behavior is likely due to the anisotropic interfacial energy of the trigonal Al₄C₃ phase [37]. When precipitates possess a more symmetric crystal structure, the restriction effect of dopant elements on precipitate coarsening is more isotropic. For example, Liu et al. [38] observed that Cu precipitates with a relatively equiaxed shape formed in a Cu-Ni-bearing low alloy steel doped with different concentrations of Al, and the size of these precipitates decreased from ~8-15 nm to <7 nm when the Al concentration increased from 0.02 wt.% to 0.98 wt.%. Later, density functional theory calculations showed that the smaller precipitate size with higher dopant concentration was due to

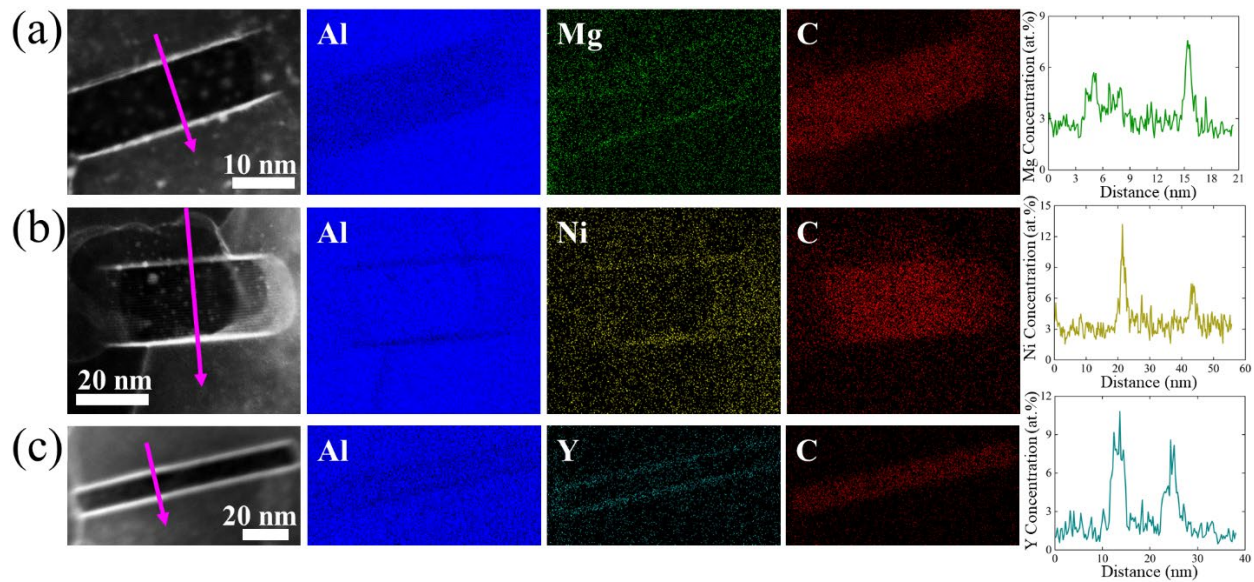

**Figure 8.** HAADF-STEM micrographs, corresponding elemental mapping, and line scans across representative nanorod precipitates in *fired + quenched* (a) Al-Mg, (b) Al-Ni, and (c) Al-Y system. Dopant atoms segregated to the longer edge of nanorod precipitates in all three alloys.

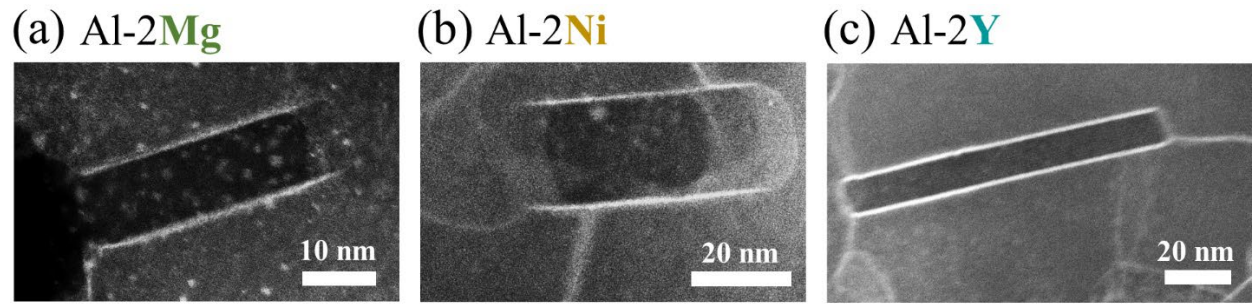

**(a) Al-2Mg**  **(b) Al-2Ni**  **(c) Al-2Y**

**Figure 9.** HAADF-STEM micrographs of representative nanorod precipitates in *fired + quenched* (a) Al-Mg, (b) Al-Ni, and (c) Al-Y system showing very little segregation to the shorter edges in all systems.

the segregation of Al to the matrix-precipitate interface, which was driven by favorable chemical interactions at the interface and resulted in a decrease in interfacial energy [39]. In the present study, because of the weak segregation to the shorter precipitate edges, the precipitate lengthening is likely less restricted than widening, resulting in the elongated nanorod shape. These observations demonstrate that precipitate morphology can be modified by controlling dopant segregation to different regions of the precipitates.

To quantify dopant segregation at the longer nanorod precipitate edges, line scans were performed and are presented in the last column of Figure 8. First, it is notable that segregation to the precipitate edges is not homogeneous, possibly due to variations in the interfacial structure between the precipitate interior and the matrix [17]. At least 16 precipitate edges were measured for each system in order to obtain more statistically reliable data, giving average nanorod interfacial compositions of $6.0 \pm 1.0$ at.% for Al-Mg, $11.2 \pm 3.7$ at.% for Al-Ni, and $12.0 \pm 3.2$ at.% for Al-Y. Therefore, the dopant concentrations at precipitate edges are similar for the Al-Ni and Al-Y alloys, being almost twice that observed in Al-Mg. With the lower level of segregation in Al-Mg, it is interesting that the precipitate width in the former system is not the largest. Thus, in addition to the segregation level, the dopant elements must have different intrinsic abilities to pin the precipitate edges. Atomic size can serve as one such metric, with large size differences with the matrix often predicted to have a strong stabilizing effect [40,41]. With this in mind, Y would be expected to show the highest stabilizing effect, followed by Mg and then Ni. This order is consistent with the trends in precipitate width, where Al-Y exhibits the narrowest precipitates with an average value of 7.4 ± 2.4 nm, smaller than that of Al-Mg (10.2 ± 3.5 nm), while Al-Ni shows the largest precipitate width of 15.1 ± 3.9 nm.

Figure 10 shows a compilation of dopant concentration at all examined grain boundaries (labelled as "GB") and nanorod edges (labeled as "NE") as well as the corresponding average

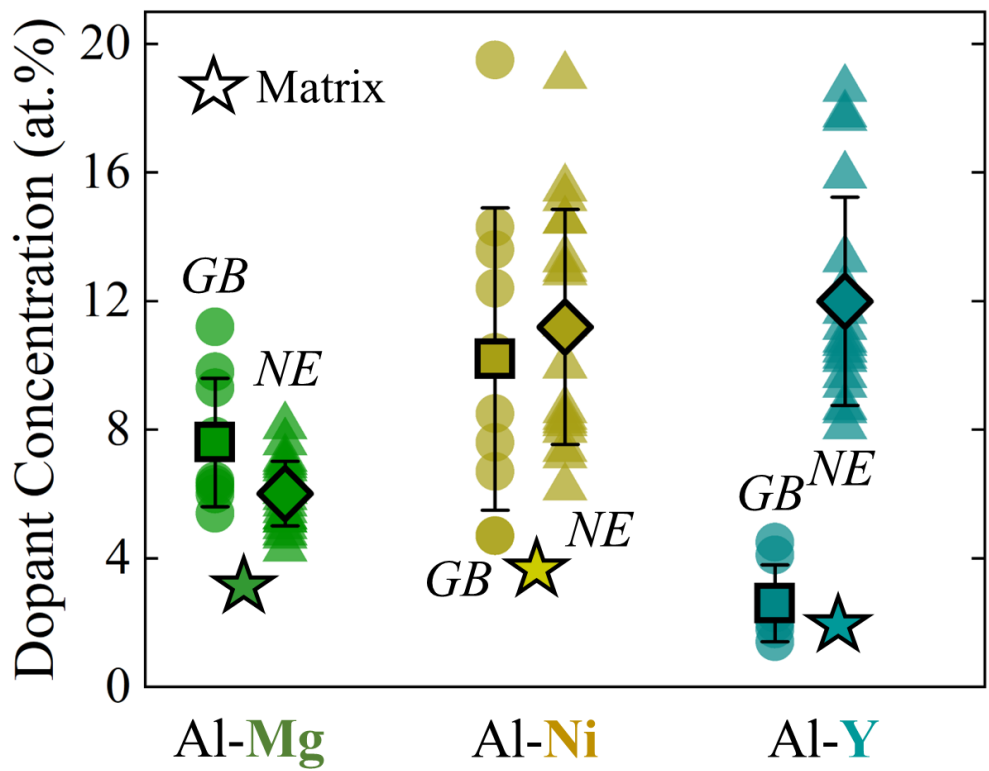

**Figure 10.** A compilation of dopant concentration at grain boundaries (labelled as "GB") and nanorod edges (labelled as "NE") for the *fired + quenched* Al-Mg, Al-Ni, and Al-Y alloys. The matrix concentration is denoted as a star with the corresponding color for each system.

can be lowered.

values. Since no obvious difference was observed for boundaries with and without nanorod precipitates (see Figures 5-7), all boundaries are shown together in this figure without further distinction. For the Al-Mg alloy, the concentration at grain boundaries (7.6 ± 2.0 at.%) is higher than at nanorod edges (6.0 ± 1.0 at.%), while these segregation levels are similar (10.2 ± 4.7 at.% for GB vs. 11.2 ± 3.7 at.% for NE) in the Al-Ni alloy. However, the Y concentration at nanorod edges (12.0 ± 3.2 at.%) is much higher than at grain boundaries (2.6 ± 1.2 at.%). The different behavior across the three systems is most likely related to the density of nanorod precipitates. A high density of nanorods in the microstructure, as in the case of Al-Y, means that dopant atoms can be pulled from the grain boundaries into the nanorod edges and grain boundary concentration can be lowered.

Recalling that Figure 2 shows that Al-Mg and Al-Ni experience abnormal grain growth while the grains in Al-Y remain nanosized, a combination of a large amount of heavily doped nanorods, modest grain boundary segregation, and plentiful amorphous complexions can stabilize a nanocrystalline grain structure. Therefore, amorphous complexions and nanorod precipitates appear to be the key factors for avoiding abnormal grain growth in such hierarchical nanostructured alloys, as these features compensated for the relatively limited amount of grain boundary segregation in Al-Y. It is worth emphasizing that the grain size of Al-Y is close to that of a ternary Al-Ni-Y system with the same processing condition. Therefore, this binary system serves as an example of a simpler alloy where fewer dopant elements can be used to achieve similar properties as multi-dopant systems through engineering grain boundaries, which may be an important strategy to improve the sustainability of structural materials (see, e.g., [42]).

In the Al-Ni and Al-Y alloys, some dopant atoms also participate in the formation of $Al_3Ni$ and $Al_3Y$ particles. Detailed examination of the intermetallic phase revealed that nanorods were always found within the particles, and Figure 11 presents two examples for each alloy system. The morphology of these carbides is the same as those formed at the grain boundaries between matrix grains (see Figures 8(b) and (c)), where the precipitates in Al-Ni are close to equiaxed and those in Al-Y have a high length-to-width aspect ratio. The coexistence of these features leads to the conclusion that the intermetallic phases grow from the enriched edges of the nanorod carbides, and transport of dopants from the grain boundary network, out along the nanorod edges, feeds the growth of the intermetallics. For carbon nanotube reinforced Al alloys, it has been shown that the $Al_4C_3$ may form when the processing temperature is above 723 K [43], which is due to a reaction of amorphous C on the nanotube surface with Al at high temperatures [44]. Therefore, these carbides were preferentially at the residual amorphous carbon layers on the surface of the carbon nanotubes [43,44].

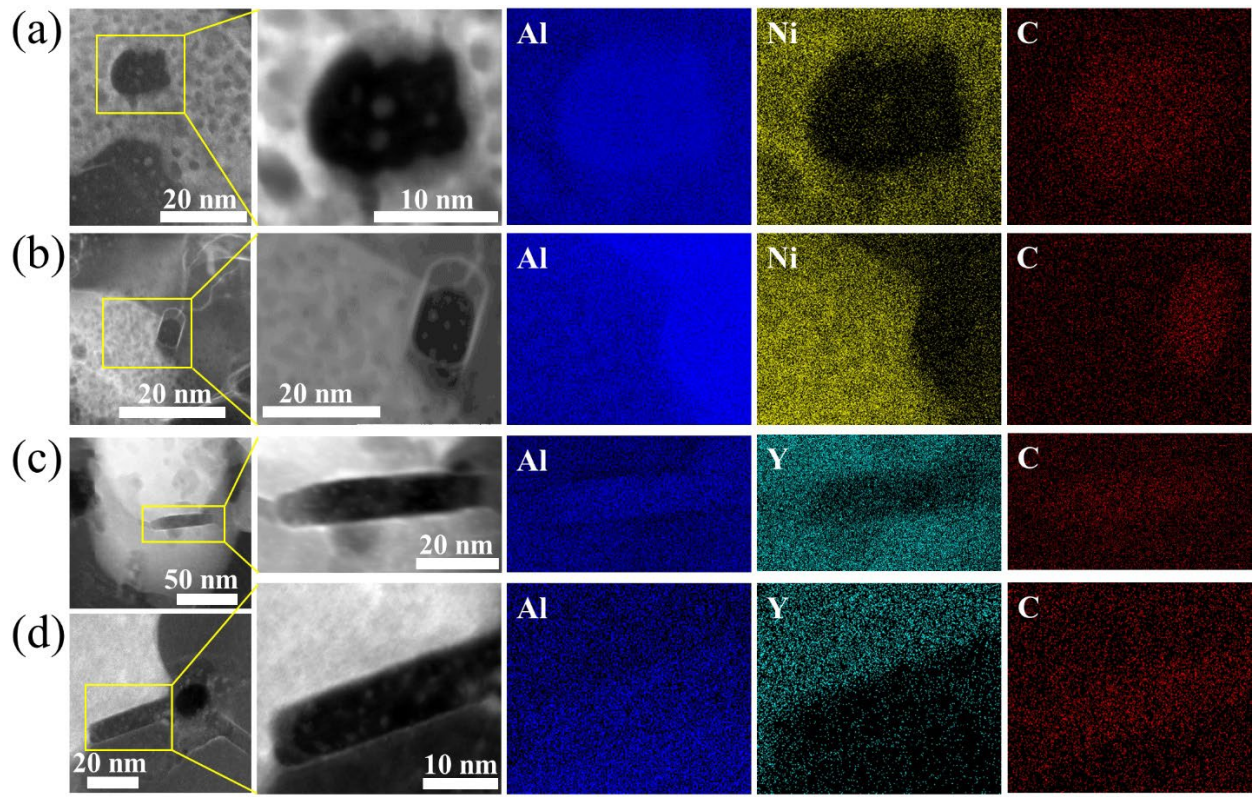

**Figure 11.** Low-mag and higher-mag HAADF-STEM micrographs as well as the corresponding elemental mapping for nanorod precipitates within intermetallic phases in *fired + quenched* (a), (b) Al-Ni, and (c), (d) Al-Y.

All three alloy systems exhibit a heterogeneous microstructure that originates from the segregation of dopants to grain boundaries. Instead of staying within the boundaries, these dopant atoms facilitate the nucleation and formation of various microstructural features through a dopant diffusion process demonstrated in Figure 12 that is uncovered in this study by the detailed investigation of interfacial chemistry presented in Figures 5-10. As the temperature increased from the room temperature to the target hot-pressing temperature, the dopant elements can first segregate to grain boundaries at relatively low temperature. For example, Koju and Mishin [45] studied Mg segregation and grain boundary diffusion in an Al-5.5 at.% Mg alloy by combining molecular dynamics and Monte Carlo methods, observing that the local concentration of Mg at grain boundaries can exceed 20 at.% at a temperature of 350 K, or approximately $T_H$ = 0.37. As $T_H$ reached ~0.65 [25], the heavily doped grain boundaries would transform to a disordered structure in order to lower their free energy, forming amorphous complexions. Garg et al. [33] investigated the interfacial segregation and complexion transitions in a Cu-Zr alloy at two different temperatures, 900 K ($T_H$ ~0.66) and 1050 K ($T_H$ ~0.77), using Hybrid Monte

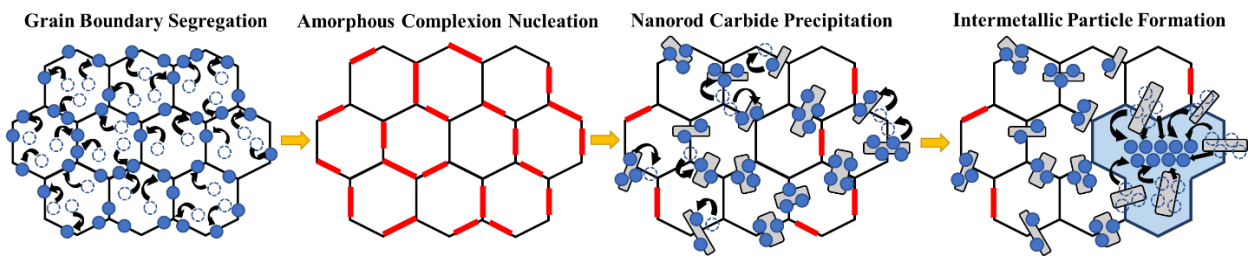

**Figure 12.** Schematic illustration of dopant diffusion pathways and microstructural evolution during the materials fabrication process.

Carlo/molecular dynamic simulations. At 900 K, only 6% of the boundaries transformed to amorphous complexions, while the fraction corresponding to the case of 1050 K was increased to 23%, pointing to the positive effect of higher temperatures on the formation of more amorphous complexions. Following amorphous complexion formation, nanorod carbide precipitates can nucleate and grow by absorbing dopant atoms from these transitioned grain boundaries. It is worth noting that unlike most grain boundary precipitates which have a uniform chemistry [16,46], the carbide precipitates in the present study possess a core-shell structure with the shell being enriched with the dopant atoms. Consequently, the precipitate morphology was significantly affected by the dopant segregation to the precipitate edges. Finally, intermetallic phases can nucleate from the dopant-enriched regions along the edge of the nanorod precipitates and create the final hierarchical microstructure. The commonality of the microstructural evolution pathway for the three alloys is their ability to form amorphous complexions that act as nucleation sites for other microstructural features, with the characteristics of these complexions (e.g., thickness and volume fraction) controlling the distribution of the other features. Most notably, larger complexion thicknesses and therefore a larger fraction of amorphous boundaries give rise to a higher number density of carbide precipitates. These collective effects likely underpin the temperature-dependent mechanical behavior in the related ternary alloys, as well as their tailorability with alloying [47].

## 4. Summary and Conclusions

In the present study, three binary Al-based alloy systems (Al-Mg, Al-Ni, and Al-Y) were examined to investigate the amorphous complexion stability, segregation competition between various interfacial features, and the dopant diffusion pathways for microstructural evolution within heterogeneous nanostructured materials. The development of the complex microstructure had important implications for grain size stability.  The stability of the Al-Y alloy is the highest, since all grain sizes were below 100 nm for both conditions, while abnormal grain growth occurred in the Al-Mg and Al-Ni alloy systems.  The following important conclusions are drawn:

(1) Y is the most efficient amorphous complexion stabilizer, as these features were observed in both n*aturally cooled* (very slowly cooled) and *fired + quenched* (rapidly cooled) samples.  For the Al-Ni system, amorphous complexions were retained only after fast quenching because of the high cooling rate, while Al-Mg did not contain amorphous complexions following any cooling conditions.

(2) Nanorod precipitates formed at grain boundaries in all three systems but showed different number density and morphology depending on the chosen dopant element.  Al-Y possessed a much higher number density and a larger length-to-width aspect ratio than Al-Ni and Al-Mg, most likely due to a higher fraction of amorphous complexions within the grain boundary network and the stronger pinning effect of large Y atoms on the longer precipitate edges, respectively.

(3) Although the grain boundary dopant concentrations were highest in Al-Ni, followed by Al-Mg and finally Al-Y, the enrichment factors at these grain boundaries (defined as the ratio of dopant concentration at grain boundaries to the matrix) were similar across the three systems.  For precipitate edges, the dopant concentrations and enrichment factors were highest for Al-Y, followed by Al-Ni and finally Al-Mg.  Of the three dopant element choices, Y was an efficient additive as most of the atoms segregated to either grain boundaries or nanorod edges and very few remained within the matrix, with segregation to the nanorod precipitates being more important for grain size stability than grain boundary segregation for this set of alloys.

(4) Nanorod carbide precipitates that nucleated and grew from the amorphous complexions were also observed within all secondary intermetallic particles, suggesting a diffusion pathway of the dopant atoms during the fabrication process. Dopants first segregated to grain boundaries at low temperatures, which then transitioned to amorphous grain boundaries as the temperature rose above a critical threshold value. Subsequently, nanorod precipitates nucleated and grew at grain boundaries by absorbing dopants from the amorphous complexions and the connected grain boundary network. Finally, intermetallic particles formed in the dopant-enriched regions containing nanorods, resulting in a complex, heterogeneous microstructure.

(5) The combined effect of being an efficient complexion stabilizer and a dopant that segregates very strongly to the nanorod edges gave the Al-Y alloys the highest resistance to grain growth, as the only alloy which did not experience abnormal growth at the very high temperatures used for processing and subsequent annealing.

As a whole, the present study provides valuable insights to how various dopant species affect interfacial structural transitions and the development of heterogeneous nanostructures. The findings shed light on how segregation engineering can be used as a tool for the design of nanostructured alloys with stable microstructures through the manipulation of interfacial features, and ultimately improved properties.

**Declaration of Competing Interest**

The authors declare that they have no known competing financial interest or personal relationships that could have appeared to influence the work reported in this paper.

**Acknowledgements**

This research was supported by the U.S. Department of Energy, Office of Science, Basic Energy Sciences, under Award No. DE-SC0021224. The authors acknowledge the use of facilities and instrumentation at the UC Irvine Materials Research Institute (IMRI), which is supported in part by the National Science Foundation through the UC Irvine Materials Research Science and Engineering Center (DMR-2011967). SEM, FIB, and EDS work was performed using instrumentation funded in part by the National Science Foundation Center for Chemistry at the Space-Time Limit (CHE-0802913).

# Supplemental Materials

<u>Section 1: Calculation of Y distribution between different microstructural features</u>

A back-of-the-envelope calculation of the dopant concentration for the *fired and quenched* Al-Y system is detailed below, which suggests that the low concentration of Y in the matrix grains and at grain boundaries is due to the segregation to the edges of nanorod precipitates. From the XRD analysis, the weight percent of each phase in the *fired and quenched* Al-Y alloy is 81.7 wt.% for Al, 8.1 wt.% for Al$_3$Y, and 10.2 wt.% for Al$_4$C$_3$ (Table I). These weight percent values can be converted to volume fractions based on the density of each phase (2.7 g/cm$^3$ for Al, 3.63 g/cm$^3$ for Al$_3$Y, and 2.99 g/cm$^3$ for Al$_4$C$_3$ [1]. The calculated volume fractions of Al, Al$_3$Y, and Al$_4$C$_3$ are 84.3 vol%, 6.2 vol.%, and 9.5 vol.%, respectively.

By applying the rule of mixtures, the total dopant concentration within the sample, $C_{total}$ can be expressed as:

$$C_{total} = C_{Y\_matrix} \times V_{matrix} + C_{Y\_Al3Y} \times V_{Al3Y} + C_{Y\_Al4C3} \times V_{Al4C3} \times V_{LongerEdge}, \qquad (1)$$

where $C_{Y\_matrix}$ is the dopant concentration within the matrix (1.1 at.% from STEM-EDS data), and $V_{matrix}$ is the volume fraction of the Al matrix (84.3%). $C_{Y\_Al3Y}$ represents the Y concentration in the Al$_3$Y phase (25 at.%) and $V_{Al3Y}$ corresponds to the volume fraction of the intermetallic phase (6.2%). $C_{Y\_Al4C3}$ is the dopant concentration at the precipitate edges, the average value of which is 12.5 at.% (Fig. 10), and $V_{Al4C3}$ is the volume fraction of the precipitates (9.5%). Therefore, the only unknown parameter in Eqn. (1) is $V_{LongerEdge}$, corresponding to the volume percent of the longer edges to the precipitates.

Based on the STEM images and EDS results, the nanorods can be assumed to have a cuboid morphology with an average width of 7.4 nm as shown in the figure below. Assuming the thickness of each longer edge to be ~0.5 nm, then the volume percent of the longer edges to the precipitates, $V_{LongerEdge}$, can be estimated as (0.5 nm×2)/(7.4 nm). Plugging all values into Eqn. (1) results in $C_{total}$ = 2.6 at.%, which is close to the nominal dopant concentration (2 at.%).

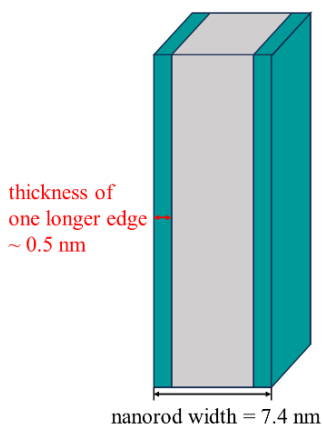

thickness of
one longer edge
~ 0.5 nm

nanorod width = 7.4 nm

**Figure S1. Schematic illustration of nanorod morphology.**